\documentclass[12pt]{article}
\usepackage[utf8]{inputenc}
\usepackage{url}
\usepackage{amsmath}
\usepackage{amssymb}

\begin{document}
\small Version vom 22. 6. 2016 \normalsize

\begin{center}\Large{\bf General relativity and the growth of a
    sub-discipline ``gravitation'' in the German speaking physics
    community}\end{center} \normalsize \begin{center} Hubert Goenner\\
  Institute for Theoretical Physics\\ University 
  G\"ottingen\\ Friedrich-Hund-Platz 1\\37077  G\"ottingen\\
 \end{center}
\section{Introduction}
During the 19th century, the topic of gravitation had been firmly
embedded into the teaching of {\em Newtonian mechanics} although,
occasionally, other theories of gravitation had also been discussed
during this century (Huygens, Secchi, Lesage-Thomson, Weber-Tisserand
etc) \cite{Isenkra1879}, \cite{Marcus1931}. Gravitational theory as a
topic of research was the exception, though; only its applications
played a role. Focal points were celestial mechanics and the study of
the  Earth's gravitational field -- in the framework of the Newtonian
gravitational force. Cf. the section ``Universal gravitation'' of 1908
in  Winkelmann's ``Handbuch der Physik'' by Felix Auerbach (1856-1933)
\cite{Wink1908} who in 1921 also contributed a popularizing booklet on
relativity theory \cite{Auer1921}. It seems interesting that even in
the 1960s, in an encyclopedic dictionary, a rigorous separation
between the entries for gravitation/gravity (with no mention of
Einstein's  theory) and general relativity was upheld
\cite{Thew1961}.\\   

Here, we will follow the growth of gravitational theory,
i.e. relativistic theories of gravitation, mainly Einstein's, as a
branch of physics in the sense of social, more precisely {\em institutional} history.\footnote{We are aware that there is an
overlap between work on {\em special} relativistic theories
and past research in general relativity.} Thus, the accompanying
{\em conceptional} development is touched only as a way of loosening
the narration and as help for a better
understanding.\footnote{Usually, physicists are more interested in the
  evolution of theoretical concepts and in observational progress
  \cite{EhSchae1990}.} As a first step, we will ask when and to what
degree {\em general relativity} became a subject of physics research
and what its relation to the ``mainstream'' in physics was. As
criteria for a continued professionalizing of the field, we use the 
numbers of published research papers, reviews, monographs, courses at
universities and possibly, appointments of lecturers or professors as 
well as the status of those involved (members of academies etc). The
foundation of special institutes for gravitational research and of
particular divisions of physical societies forms the final stage of
this process of institutionalization. With the beginning of
relativistic astrophysics in the 1960s, research work concerning
general relativity also was carried through in this new field. We will
try to find out whether there was a slump of research in general
relativity between 1925 and 1955 in Germany. For several reasons, in
the 1960s activity increased there as it also did in the United States
and elsewhere. In Germany, from the 1970s to the 1990s, a continued,
stable production of PhDs in the field of relativistic gravitation did
occur.\footnote{The time-period of this survey will end with the
  beginning of the 1990s.}\\   
  
Occasionally, there is a problem of discrimination between
contributions made by scientists from German speaking countries and
those by scientists from other countries presented while in
Germany/Austria as guest professors etc. The latter are included here
because their work is taken as sign of activity in the German speaking 
countries. Similarly, work done at the German University in Prague
(Karls-Universität), a German-language university, which existed from 
1348 to 1945, will be included. Although the Netherlands do not belong
to the set of countries selected, a few remarks must be made on early
contributions to general relativity from there.\\   

Obviously, any institutionalizing depends on the amount of financial
means. In the first half of the 20th century, the budget of
universities in the German speaking countries normally would have been
able to only support a full professorship with one or at most two
assistants (doctoral students or post-docs) interested in this new
field. A larger group of researchers would have required third party 
financing. Until the 1920s, such ``third parties'' mainly had been the
few scientific academies. In 1920, the ``Notgemeinschaft der deutschen
Wissenschaft'' was established, a self-administrative organization
combining private (industrial) and government financing. In 1929, it
became renamed (Deutsche) Forschungsgemeinschaft. Einstein did obtain
funds from it for collaborators. Another funding agency, which could be
and in 1924 was tapped by Albert Einstein, is the Rockefeller Foundation.

\section{Before World War II: Research and teaching by single 
scientists}  

\subsection{The early years of general relativity} 
\subsubsection{Progress during World War I: 1915 to 1918}
After a long maturing period, general relativity was completed in
November 1915 with field equations suggested by the physicist Albert
Einstein \cite{Einstein1915} and the mathematician David Hilbert
\cite{Hilbert1916}. (For the priority debate, cf. \cite{EarGly1978}, \cite{CoReSta1997}, \cite{ReSa1999}, \cite{Win2004}, \cite{Wue2007}.) Unlike in the case of special relativity, the physicists around Einstein in Berlin were slow in accepting the new theory. In a letter to Arnold Sommerfeld, he expressed perfect
satisfaction about his achievement, but held: ``[..] but none of the
peers has recognized up to now the depth and necessity of this path.''
In particular, Max Planck and Max von Laue were not open to his
considerations of principle \cite{ECP1915}. Nevertheless, already in
1916 and 1917 three of the best known and most useful exact solutions
of Einstein's field equations were found. The first, describing an
isolated finite spherically symmetric mass-distribution at rest, was
communicated to Einstein already in December 1915 by his colleague in Berlin, the astronomer Karl Schwarzschild (1873-1916), then in
Russia with the German army. The result was published in 1916
\cite{Schwarz1915}, \cite{Schwarz1916b}, \cite{Schwarz1916a}. In the
same year, a student of  Hendrik Antoon Lorentz in Leiden, Johannes
Droste (1886-1963), in his dissertation independently presented this
same (Schwarzschild) solution \cite{Droste1917}.\footnote{For previous work by Droste on Einstein’s theory of 1913 cf. \cite{Droste1915a}, \cite{Droste1915b}.} The aircraft-designer and professor at the Technical University in Berlin-Charlottenburg, Hans Reissner (1874-1967), followed with an exact solution for a time-independent, isolated, {\em electrically charged} spherically symmetric mass
 \cite{Reissner1916}. It was rediscovered in 1918 by Gunnar
 Nordström \cite{Nord1918}. The third important solution, the
 de Sitter-solution, is described further below. Einstein himself
 propagated his new gravitational theory in an article
 \cite{Einstein1916a} and in a book for everbody (with some physics education)
 \cite{Einstein1917a}; in a number of papers he also worked out
 consequences of the theory, among others for cosmology and
 gravitational waves  \cite{Einstein1918}. Also in 1916, the
 astronomer Erwin Freundlich (1885-1964) of the Potsdam observatory,
 who would become Einstein's payed collaborator in 1918, published a
 booklet on general relativity \cite{Freundlich1916} to which Einstein
 wrote a preface backing the author:\begin{quote}``I have gained the
   impression in perusing these pages that the author has succeeded in
   rendering the fundamental ideas of the theory accessible to all who
   are to some extent conversant with the methods of reasoning of the exact
  sciences. (English translation of 1922 by H. L. Brose
  \cite{Freundlich1922}.)\end{quote}  Freundlich had been interned in
1914 in Russia, after World War I had broken out, during a
solar-eclipse expedition. The physicist turned philosopher Moritz
Schlick (1882-1936) who for two years fulfilled his military service
at an air-base in Berlin, apparently had time to write a book, liked by Einstein, on philosophical consequences of relativity theory published in 1917 \cite{Schlick1917}. \\  

In Göttingen, the mathematicians David Hilbert and Felix Klein looked more closely into the mathematical structure of Einstein's equations and some simple solutions. Hilbert again derived the Schwarzschild metric \cite{Hilbert1917}, and Klein pointed to the possibility of an elliptic
geometry of constant curvature in place of the spherical one. He
convinced Einstein that de Sitter's solution is free of singularities,
and investigated conservation laws for energy and momentum
\cite{Klein1918a}, \cite{Klein1918b}, \cite{Klein1918c}. From the
mathematical side, a most important contribution appeared in 1918: the
book by the mathematician Hermann Weyl (1885-1955) \cite{Weyl1918}, who in 1908 had obtained his PhD with D. Hilbert in Göttingen and at the time was professor at the ETH Zürich. Weyl also found the exact
solution to Einstein's equations with cosmological constant (free of matter)
corresponding to the Schwarzschild solution \cite{Weyl1919}. It was
rediscovered in 1922 by the mathematician Erich Trefftz (1888-1937)
\cite{Treff1922}. Very early, in 1916, the mathematician Gustav
Herglotz (1881-1953), formerly in Göttingen and then in Leipzig, also 
published on Riemannian geometry and Einstein's gravitational theory
\cite{Herglo1916}.\\

In Vienna, quite a few physicists and mathematician had followed
Einstein's work already before its culmination in 1915
\cite{Havas1999}. Best known among them are Friedrich Kottler
(1886-1965), Hans Thirring (1888-1976), Josef Lense (1890-1985), Ludwig
Flamm (1885-1964) and Erwin Schrödinger (1887-1961). Kottler had
critically reacted to general relativity \cite{Kottler1916} and had
received an answer by Einstein, ``[..] because this colleague really
has grasped the spirit of the theory'' \cite{Einstein1916b}. Kottler
became an unpaid lecturer in Vienna and thanked Einstein ``for the benevolence toward me which you kept''\cite{Kott1920}. In 1922,
he independently re-derived the solution found by Weyl and \linebreak
Trefftz and concluded that Maxwell's equations could be formulated without the use of either metric or  connection \cite{Kottler1922a}. Flamm introduced a pictorial representation of the Schwarzschild geometry which is used even today for picturing black holes \cite{Flamm1916}. The first approximative solution of the gravitational field for a (slowly) 
{\em rotating} solid body was given in 1918 by the astronomer, and then mathematician Josef Lense (1890-1985) and the physicist Hans Thirring \cite{LenThi1918}; the
perturbation of all orbital elements were calculated. Thirring himself
had worked on a rotating mass before: for a hollow rotating sphere, he
derived Coriolis and centrifugal forces \cite{Thirring1917a}. In July
1917, he wrote to Einstein that: ``[..] the young Viennese school is
occupied intensely with gravitational theory.'' He hoped that Einstein
would give them further advice, because since the death of Hasenöhrl
they were on their own \cite{Thirring1917b}. (For english translations
and a thorough discussion of  the papers by Thirring and Thirring \&
Lense cf. \cite{BaHeThe1984}.) In a brief note, Lense then published
the calculated effects due to the rotation of the Earth and outer
planets for some of their moons \cite{Lense1918}. Another Viennese,
Hans Bauer (1891-1953), in 1918 investigated systems of spherically
symmetric fluids with a linear equation of state \cite{Bauer1918}, for
which the simplest solution already had been given by
K. Schwarzschild. Erwin Schrödinger \cite{Schroed1918a} calculated the
components of Einstein's energy-momentum complex for the external
Schwarzschild metric, found that all vanish in particular coordinates,
and consequently  doubted its physical importance. Even more
devastating was Hans Bauer's paper showing that Minkowski space in
particular coordinates led to non-vanishing components of the suggested energy density of the gravitational field \cite{Bauer1918a}. In
another paper, Schrödinger showed that the Einstein cosmos, found by Einstein as an exact solution of his field equations {\em with}
cosmological constant, could be obtained from the original field
equations ({\em without} cosmological constant) by a slight generalization
of the fluid matter in the matter tensor \cite{Schroed1918b}.\\ 

At the time, Prague with its Karls-Universität still belonged to the
Habsburg (Austrian-Hungarian) Empire. There, since 1912 until 1938,
Philipp Frank (1984-1938) was succeeding Albert Einstein on his
professorship. He occasionally argued against philosophers on subjects
connected with special relativity \cite{Frank1937}, but is best known
by his book on Einstein \cite{Frank1947}. Until the Nazi takeover he
tutored Jewish students from Germany (cf. below).\\  

In Switzerland, apart from Hermann Weyl in Zürich, nobody followed
Einstein's achievement of 1915 with serious research-work on general  
relativity, not even Marcel Grossmann (1878-1936) who had contributed
much to the mathematical formulation of Einstein's theory
\cite{EinGros1913}, \cite{EinGros1914}, \cite{Sauer2013}. Paul Gruner
(1869-1957) of the University of Bern worked on the kinematics of
special relativity; a presentation of general relativity is included 
only in his booklet of 1922 \cite {Gruner1922}. Einstein's former
colleague at the patent office, Edouard Guillaume (1881-1959), since 1917 in correspondence pestered him with queries and own
principles.\footnote{As one example cf. \cite{Guillau1920}.}\\   

Outside of the German speaking community, but in close contact with
Einstein, others with an active interest in general relativity
must be mentioned: the circle in Leiden around the Dutch theoretical
physicist and Nobel prize-winner Hendrik A. Lorentz (1853-1928) and
his colleague, mathematician and astronomer Willem de Sitter
(1872-1934). In March 1917, de Sitter presented the first exact solution of
Einstein's field equation {\em with cosmological constant}, without
matter, describing the geometry of a 4-dimensional space of constant,
positive curvature \cite{deSit1916}. It was to play an important role
for cosmology. From de Sitter's publications during the first world war, English
scientists learned about general relativity \cite{deSit1916a},
\cite{deSit1916b}, \cite{deSit1917a}. For Lorentz, Droste and de
Sitter see the articles by A. J. Kox \cite{Kox1993}, \cite{Kox1992}.\\   

It is surprising that already during World War I and before its end
in November 1918, such a sizeable number of contributions to general
relativity could have surfaced.\footnote{How World War I influenced physicists in Germany is described in \cite{Wolff2002}.} With the exception of H. Reissner, who
had passed the of age forty, the others in Berlin and Vienna were drafted
into military service; Flamm and Thirring had been dispensed, or
allowed to do research valuable for the military in
Austria. Schwarzschild died due to an autoimmune skin disease which
broke out during his service at the front; also the Viennese physicist
Max Behacker (1885-1915) who had worked on Nordström's
gravitational theory was killed in action. In fact, the
Nobel prize-winner Wilhelm Wien (1864-1928), ``on the instigation of
the supreme command of the eighth army'', held lectures in the Baltic
states in which he already discussed Einstein's new gravitational theory
\cite{Wien1918}.\\ 
An exceptional position was taken by a distinguished scientist remaining critical with regard to general relativity and suggesting an extended gravitational theory of his own: Gustav Mie (1868-1957), first in Halle and then, from 1924, in Freiburg/Breisgau \cite{Mie1921}, \cite{Koh2002}. There, research in general relativity was resumed after World War II.\\ 
Outside of academia, the teacher at a Gymnasium in Wilhelmshaven,
Ernst Reichenb\"acher (1881-1944) made a very early attempt at a
unified field theory of gravitation, electromagnetism, and
matter\footnote {Reichenb\"acher studied mathematics and in 1903 received his doctorate
  from the University of Halle. At first, he did not enter an academic
  career, but started teaching at a Gymnasium in Wilhelmshaven at the North Sea, then in K\"onigsberg on the Baltic Sea. In 1929 he
  became a Privatdozent (lecturer) at the University of K\"onigsberg
  (now Kaliningrad, Russia). His courses covered special and general
  relativity, the physics of fixed stars and galaxies with a touch on
  cosmology, and quantum mechanics. In the fifth year of World War II
  he finally received the title of professor at the University
  K\"onigsberg, but in the same year was killed during a bombing raid
  on the city.} \cite{Reichen1917a}, \cite{Reichen1917b}. \\

Thus, in the German speaking countries, from 1915 to 1918 research on Einstein's new theory of gravitation thus was done mainly at the
Prussian Academy of Sciences and at universities in Berlin, Göttingen and Vienna around Einstein, Hilbert, Klein, and Hasenöhrl - also after his early
death. Another germ for this kind of research formed in the neutral
Netherlands at the University of Leiden around H. A. Lorentz and the
mathematician Jan Arnoldus Schouten (1883-1971) in Delft
\cite{Schout1919}, \cite{Schout1924}. In the warring countries opposed
to Germany, the reception of general relativity was much slower. Exact
solutions of Einstein's field equations could have been derived by
anyone having some knowledge about partial differential equations
without  need to understand the physics behind them. My impression is that, beyond those more mathematically interested physicists doing
research in the field of general relativity during that time, all
belonged to a small network close to Einstein, centered in Germany, and without any other ``institutional structure'' than the Prussian
Academy of Sciences and his Kaiser-Wilhelm-Institute for Physics
opened in 1917. While, on particular request by Einstein, the Academy
could pay a coworker, the budget of his Physics Institute could be
spent only in part for research in gravitation \cite{GoeCas2004}. The
activity in the field of general relativity was as constrained as that
Einstein could keep track of  all what went on. From his
correspondence we learn that he knew about everyone working on his
gravitational theory; he reacted defensively to publications pointing
out some weakness in his formalism,
cf. \cite{Einstein1918a}, \cite{Einstein1918b}. \\      

Einstein had also close ties with the editor of the scientific
magazine ``Die Naturwissenschaften'' built after the corresponding
English journal ``Nature'', Arnold Berliner (1862-1942), who used to
ask him for his advice (cf. \cite{Rowe2006}).    

\subsection{The media hype: 1920-1924}
The announcement in 1919, by two British solar-eclipse-expeditions on the verification of Einstein's prediction concerning the light deflection 
by the Sun, set off an avalanche of public reactions in Germany. One of
them was the unprecedented scientific glorification of Albert
Einstein, the new Archimedes, Kopernikus, and Newton in one. Another,
the flood of publications, both pro and con general relativity, in the
form of popular brochures and serious books. Its volume and
distribution over the years may be taken from my article of 1992
\cite{Goenner1992}. As discussed there, a correlation with the
financial and economical situation in Germany is likely. There are
indications that Einstein was more than willing to encourage those who
were spreading the word about his theories. An example for this is
given by Moszkowski's book ``[..] on relativity theory and a new
system of the world modeled from conversations with Einstein''
(\cite{Mosz1921}, subtitle).\\   
  
We will mention just a few examples for both scientific and popular
publications during this period. After his successful monograph on
special relativity \cite{Laue1911}, Max v. Laue wrote a second part
``On general relativity and Einstein's teachings on gravity''
\cite{vLaue1922}. Unlike v. Laue's rather technical book, Hans Thirring
published one without any formula: with the explicit aim ``to uncover
the relations among the basic ideas of relativity theory''
\cite{Thirring1921}. Likewise, Max Born wrote an
``easy-to-understand'' book on relativity theory, not as heavy as
v. Laue's, but still with many formulas \cite{Born1920}. It was
successful; in its third edition, Born omitted the picture of Einstein
on the frontispiece ``[..] in order to avoid the appearance that
personal sympathy mixes with my scientific convictions [...]''
(\cite{Born1922}, p. IX ). Wilhelm Wien, in a lecture of March 1921
given at Siemens \& Halske in Berlin, tried to describe both special
and general relativity ``sine ira et studio''. He accepted only three
objections toward the theory: Contradictions within the mathematical
structure, consequences which are not in agreement with
experiment/observation, or unsuitability of the theory for a final
representation of natural processes due to an abandonment of simple
principles \cite{Wien1921}.\footnote{``Wirkliche Einwände gegen die
  Relativitätstheorie können daher nur von dreierlei Art
  sein. Entweder man muß innerhalb des mathematischen Systems der
  Theorie Widersprüche aufdecken. Oder man muß zeigen, daß sie zu
  Folgerungen führt, die mit der Erfahrung nicht übereinstimmen oder
  schließlich muß man den Nachweis führen, daß sie sich wegen der
  Preisgabe einfacher Grundsätze zur endgültigen Darstellung der
  Naturvorgänge nicht eignet''.}\\ 

As to brochures and booklets for the general public, Einstein
apparently was aware of quite a number of them. He had read the
manuscript of a text by his previous student in Berlin, Werner Bloch
(1890-1973), probably his book of 1918 \cite{Bloch1918}, which had
reached its 3rd edition in 1921. He told him: \begin{quote}``With
  interest, I read through your introductory work on relativity
  theory and convinced myself that your statements are thorough,
  easily comprehensible and well arranged''
  \cite{EinBlo1917}.\end{quote}  In a recommendation by him four years
later, the text was described as ``a good booklet for an introduction
into relativity theory'' \cite{EinBlo1921}. Einstein also knew
Pflüger's brochure \cite{Pflueger1920}, ``a not badly written 
popularizing text about relativity'' \cite{EinMoch1920}. Other authors
just sent their writings to Einstein, like Harry Schmidt (1894-1951)
from Hamburg whose booklet reached several German editions and was
translated into English and Italian \cite{HSchmidt1920},
\cite{Calend1921}. Einstein's adversaries blamed him for his explicit 
support for an author of a popularizing brochure who claimed in his
preface that he was no longer able to prove ``a modest theorem of
Euclidean mathematics by hand'' \cite{Hasse1920}.\\ 

Well before this overwhelming public interest in Einstein's theories,
lecture courses on general relativity had sprung up at various
universities in the German speaking countries: Already in 1918
L. Flamm lectured at the University in Vienna, while Ludwig Hopf in
Aachen and W. Matthies in Basel\footnote{In 1920, Matthies became Dean
  of the Philosophische Fakultät  (with its
  Mathematisch-naturwissenschaftliche  Abteilung) of the University of
  Basel.} gave their courses during the winter term 1918/19. In
Berlin, Einstein followed in the summer term of 1919.  Some of the
courses were then expanded into books: In Heidelberg, the astronomer
August Kopf (1882-1960) regularly lectured about relativity theory
from the winter term 1919/20 until 1923 \cite{Kopf1921}. In Marburg,
the mathematician Ernst Richard Neumann (1875-1955) followed suit in
1920/21 \cite{Neuma1922} whereas one year later, in the winter term 1921/22,
Wilhelm Lenz, then at the University of Rostock, lectured on
``Relativity Theory II (general)'' \cite{Lenz1921}. A former teacher
and insurance mathematician Paul Riebesell (1883-1950) even prepared a
booklet on relativity theory as an assistance for the teaching in
college or high school \cite{Riebesell1926}.\\

Besides Moritz Schlick, other philosophers also had discovered the
field. Most of them were in direct contact with Einstein and supported
by him: His former student Hans Reichenbach (1891-1953)
\cite{Reich1920} asked for Einstein's permission to dedicate his book
to him; eventually, he became sort of a philosophical ``watchdog'' for
Einstein; for Schlick and Reichenbach cf. \cite{Engler2007}. Ernst
Cassirer (1874-1945) \cite{Cassi1921} sent Einstein his manuscript who
read  and praised it. Ilse (Rosenthal)-Schneider (1891-1990)
\cite{Schneid1921} wrote her dissertation in Berlin and corresponded
with Einstein as did also Joseph Petzold (1862-1929)
\cite{Petz1921}. For Einstein's relation to Kantian philosophy, as
represented by  Ewald Sellien (1893- ?) \cite{Sell1919} and Alfred
C(oppel) Elsbach (1897-1932) \cite{Elsba1924} among others, see the
article by K. Hentschel \cite{Hentsch2009}. Sellien, who later became
a teacher at a Gymnasium, had dedicated a copy of his dissertation to
Einstein ``in high respect and reverence'', but the respected did not
like his work and wrote to Schlick: ``Did you see the quite silly
dissertation by Sellien (student of Riehl)?'' \cite{Eincor1919}. \\

Besides these full time philosophers, philosophically inclined
scholars with an interest in physics also published on general
relativity. An example is Franz Selety (Jeiteles) (1893-?) in Vienna who
corresponded with Einstein and criticized his cosmological
ideas. Instead, he proposed a hierarchically arranged cosmological
model within Newtonian gravitational theory \cite{Selety1922} which
was not accepted by Einstein (cf. \cite{Einstein1922} and the
response by Selety \cite{Selety1923a}). After further papers, Selety
disappeared in Paris without leaving a trace, scientific or
otherwise.\footnote{Cf. the biographical sketch by H. Kragh
  \cite{Kragh2008}.} \cite{Selety1924}.\\   

A scientific highlight in the first half of the 1920s coming from {\em
  outside} the German speaking physics community was the presentation
of time-dependent homogeneous and isotropic exact solutions of
Einsteins field equations with cosmological constant in {\it
  Zeitschrift für Physik} by the physicist and mathematician
A. Friedmann of St. Petersburg \cite{Fried1922},
\cite{Fried1924}. These solutions contradicted Einstein's
philosophical beliefs pointing to a statical universe; it took some
effort until he accepted them \cite{Einstein1922b},
\cite{Einstein1923}. 
\subsection{The Einstein Tower}
A project instigated by Freundlich and backed by Einstein was the
building of a telescope for the investigation of the solar spectrum
with high precision. Its main purpose was to measure the redshift of
the solar lines predicted by general relativity. The architect Erich
Mendelsohn (1887-1953) suggested an expressionistic building in the
form of a tower on a lengthy foundation. The telescopes inside the
tower directed the light to the measurement area in the
basement. Measurements finally could start in 1925, but did not (and
could not) agree with the predicted solar redshift until after the
1950s \cite{Hentsch1992}. Although the observatory in the
Einstein-Tower did very important work for the solar spectrum and
other topics in astrophysics, it failed completely on its intended
purpose. The building of the Einstein Tower can be interpreted as an
attempt to institutionalize general relativity within observational
astronomy, which at the time was very reluctant to include Einstein's
theory in its agenda. A successful embedding occurred only 40 years
with the newly introduced field of  relativistic astrophysics 

\subsection{Opposition against relativity theory} 
Another aspect must be taken into account: the reception of general
relativity in the German physics community was not unanimous. It was
criticized by a few influential experimental physicists like the Nobel
prize winners Philipp Lenard (1862-1947) \cite{Lena1920},
\cite{Lena1921} and Johannes Stark (1874-1957) \cite{Stark1922}, and
by Ernst Gehrcke (1878-1960), a specialist in optics at the
Physikalisch-Technische Reichsanstalt\footnote{The German national
  Metrology Institute.} in Berlin \cite{Gehr1924}. At 
first, in particular Lenard's criticism followed physical arguments
which Einstein tried to refute. However, Lenard's attitude then
transformed into pure resistance, colored by nationalism, and later
anti-semitism. Criticism was voiced also by philosophers like Hugo 
Dingler (1881-1954) \cite{Ding1921}. The arguments presented never
succeeded to convince the great majority of those working on general
relativity. Nevertheless, they might have discouraged young scientists
to enter the field.       

\subsection{The second half of the twenties}
After World War I, German researchers had been excluded from
international conferences until 1926, when Germany became a member of
the League of Nations \cite{Reinbo2013}. The boycott had weakened
since 1922 and was practically over in 1925. Einstein worked strongly
for an end of it.\\    

With the advent of quantum mechanic in 1925 to 1926, unlike with that of
general relativity, no comparable hustle and bustle went on in the
press about the new theories of Heisenberg, Born, Jordan, Schrödinger,
and Dirac. Although the public attention to general relativity
dwindled in significance, money still could be made by writing a book
on relativity. In his early report on the theory of relativity of
1920, Wolfgang Pauli (1900-1958) had included general relativity as a
single chapter \cite{Pauli1921}. With basic knowledge about general
relativity available in the meantime, it could be summed up now by its
own in encyclopedias and textbooks. Such were the contribution by
August Kopf to the volume ``Physics of the Cosmos'' of the widely used
Müller-Pouillet textbook of physics \cite{MuPou1928}, or two separate
entries in the Geiger-Scheel ``Handbook of Physics'' by Hans Thirring
and his former doctoral student Guido Beck (1903-1988)
\cite{Habu1929}. Thirring also gave a review of relativity theory in a  
newly founded scientific journal \cite{Thirring1922}.\\  

\subsubsection{How Einstein supported his coworkers financially}
From the 1920s on, Einstein had several coworkers in Berlin: Jakob
Grommer (1917-1931), Cornel Lanczos (1928-1929)\footnote{Lanczos had
  discovered a simple axially symmetric solution of Einstein's field
  equations \cite{Lancz1924}.}, Hermann Müntz 
(1928-1929), and Walther Mayer (1931-1933). In 1921, Grommer received
some money from Einstein's Kaiser-Wilhelm-Institute for Physics; in
1924 Einstein obtained a fund of 1000 \$ from the Rockefeller
Foundation. Perhaps this money was going to Grommer; it is unknown to
me who contributed his regular salary, if he had one. Lanczos had a
research fellowship from the ``Notgemeinschaft der deutschen
Wissenschaft''. Mayer was payed by the Academy of Sciences
\cite{KirTre1979}. Müntz did not need money from Einstein: Since 1924,
he was teaching mathematics at a Gymnasium in Berlin.\footnote{For a
  brief biographical entry on Müntz, see my first article on the
  history of unified field theories \cite{Goenner2004}.}   

\subsubsection{Einstein's role in the appointment of professors}
Einstein's fame brought him many requests for his advice in matters of
professorial appointments at universities. Already before he became
famed, he had effectively suggested Philipp Frank as his successor in
Prague. From Vienna he was asked about the filling of the chair the
physicist Franz S. Exner (1849-1926) would leave, in particular
whether he could recommend Felix Ehrenhaft (1879-1952)
\cite{appoint1}. In Zürich, a decision among four colleagues had to be
made; he voted for Simon Ratnovsky (1884-1945) \cite{appoint2}. In
Leipzig, a theoretical chemist was searched for the appointment as
extraordinary professor \cite{appoint3}. David Hilbert in Göttingen
wrote Einstein for his opinion on Max Born as possible (and then the
actual) successor of Peter Debye \cite{appoint4}. For the university
of Tübingen, he suggested, as possible successors of the famous
spectroscopist Friedrich Paschen (1865-1947), James Frank and Edgar
Meyer \cite{appoint5}. Walther Mayer (1887-1948) whom Einstein had
helped to gain an unpaid position as lecturer at the University of
Vienna, became his assistant in Berlin in 1929. In spite of his
growing influence in such matters, except for Lanczos I, know of no
one who obtained a professorship in the German speaking countries due
to his research in general relativity.\footnote{In fact, Lanczos never
filled his position as extraordinary professor in Frankfurt
  obtained in 1932; he remained in the United States.} At the time,
general relativity just was not deemed to have such an importance as
to be accepted as a noteworthy separate branch of physics.  
\subsubsection{Further work on general relativity}
It seems that, world wide, there was only a minor slump in scientific
publication on general relativity from the second half of the 1920s
until 1945. An inspection of the bibliographies by Combridge at King's
college \cite{Comb1965} and in Synge's book on general relativity
\cite{Synge1966} shows no remarkable change in the source-volume
refering to the period between 1915 and 1960. What happened was that
the focus of research on general relativity shifted from Germany to
other countries. After 1927, Einstein published on the problem of
motion with his assistant Jakob Grommer (1879-1933) \cite{EinGro1927},
\cite{Einstein1927}. In 1931, he applied for and obtained funds from the
Rockefeller Foundation for a visit of Myron Mathisson (1897-1940) from
Cracow to Berlin who had made progress in the study of equations of
motion of pole-dipole particles.\footnote{Tilman Sauer has given a
  precise chronology of the application to the Rockefeller Foundation
  \cite{Sauer2007}. In the end,   the visit of M. did not
  materialize. For some facts about Mathison's life, see
  \cite{SaTra2008}.} Since 1925, Einstein more and more published on
generalizations of his theory to five-dimensional or non-riemannian
spaces; others like Cornel Lanczos in Frankfurt continued with
research in general relativity \cite{Lancz1927a},
\cite{Lancz1927b}.\footnote{For Lanczos' early contributions to
  general relativity cf. \cite{Stachel2002}, pp. 449-518.}\\   

Erwin Freundlich pursued his work for the empirical support of
general relativity. In 1929, with astronomers H. von Klüber and A. von
Brunn from the Potsdam Observatory, he undertook a
solar-eclipse-expedition to Sumatra. For the light-deflection at the
brim of the sun they found a higher value than that predicted by
Einstein's theory \cite{FreKluBru1931}. \\ 

In Vienna, Flamm, Kottler and Thirring no longer
did research work on gravitation; Hans Bauer \cite{Bauer1928} and
Guido Beck still contributed \cite{Beck1926} as well as Paul
Lazarsfeld (1901-1976) with a doctoral thesis
\cite{Lazar1926}. Lazarsfeld later turned to social research; he is
called the ``founder of modern empirical sociology.'' Hans Bauer
earned his living as a teacher at a Gymnasium. Kottler wrote two
historical papers on relativity \cite{Kottler1924a},
\cite{Kottler1924b} but then went into optics
\cite{Kottler1923}. Schrödinger had been called to Zürich in 1920,
where he found his wave equation, and then to Berlin in 1927. \\     

The end of the 1920s reflects a discrepancy between the level of
awareness about Einsteins relativity theories among the general public
in Germany, and the rather restricted professional possibilities for 
doing research on Einstein's theory of gravitation. We note that the
majority of those standing out for their contributions to general
relativity – Einstein, von Laue, Pauli and the mathematician Weyl
included – had won or earned their reputation in fields of physics unrelated to
gravitation. The possibility for obtaining a full-time job within the
field of general relativity just did not exist. According to the
majority-opinion, general relativity with its prediction of only three
observable effects in urgency had fallen way behind the new quantum
physics with its many applications. We can only guess what would have
happened if quantum mechanics had {\bf not} come into being, but it
seems safe to say that research on general relativity would not have
augmented dramatically. 


\subsection{Research until and during World War II}
\subsubsection{The thirties}
During the 1920s and 1930s, due to new powerful telescopes in the
United States the hope grew that something about the large scale
distribution of matter in the cosmos could be found out. In the
context of Friedmann's solutions, independently found by the
Belgian priest and astronomer, Georges Lemaître (1894-1966)
\cite{Lema1927}, \cite{Lema1931a}, \cite{Lema1931a}, a noteworthy
development started: Einstein finally accepted the ``expanding''
solutions and, with Willem de Sitter, created the Einstein-de Sitter
cosmological model \cite{EindeS1932}; cf. also
\cite{Einstein1931}. Before him, the astronomer Otto Heckmann
(1901-1983) at the observatory in Göttingen had independently
published detailed investigations of Friedmann's solutions
\cite{Heck1931}, \cite{Heck1932}. De Sitter who had received a copy of 
Heckmann's publication mentioned Heckmann's name in his joint paper
with Einstein without giving a precise reference.\\ 

A different contribution came from the American mathematician Oswald
Veblen (1880-1960) who in 1932, as a guest professor, lectured on
projective relativity in Göttingen, Hamburg and Vienna. The lectures
were published as a book [in German language] in the following year
\cite{Veblen1933}. This research-topic found its continuation in
Germany after World War II in Hamburg and Berlin (cf. section 3.1).\\

Einstein's interest had shifted to his attempt at using geometries
more general than Riemann's for building a field theory unifying the
gravitational and electromagnetic interaction and, if possible,
including even further particles. While in 1931 he formally supported the
application to the Rockefeller Foundation for funds allowing the
construction of an as yet not existing building for the
Kaiser-Wilhelm-Institut für Physik, Max von Laue and others in the
committee were the real promoters. Gravitation does not seem to have
been envisaged as one of the main topics to be studied in the building.\\

\subsubsection{The impact of the Nazi rule}
When the National Socialists had gained power over Germany in 1933, the
number of scientists involved in general relativity decreased
considerably. Einstein and his Jewish assistant for calculations,
Walther Mayer, emigrated to Princeton. Einstein's former assistant in
1928/29, Cornel Lanczos, since 1931 as a guest professor at Purdue University in
Indiana, USA, did not return to Germany.\footnote{As Einstein stated
  explicitely that, during all of his life, he occupied only Jewish assistants.} F. Kottler and H. Reissner due to their Jewish
ancestors also had to leave the country.\footnote{Kottler earned his
  living in the optical industry in the USA while Reissner became
  associated first with the Illinois Institute of Technology and then
  with the Polytechnical Institute of Broklyn \cite{Reissner1977}. I
  owe this reference to F. W. Hehl, Cologne.} In 1933, Felix Auerbach
in Jena committed suicide together with his wife. Hans
Thirring was forced into retirement in 1938 after the  German
occupation of Austria. This also happened to Schrödinger at the
university in  Graz who regularly lectured in Vienna
(cf. \cite{Havas1999} p. 190-192). Thus, no further research on 
general relativity was done in Vienna until after World War II. In
Berlin, the situation was not much better: Max von Laue, although in
opposition to Nazi ideology, stayed on. His assistant of 1933 to
1940, Max Kohler (1911-1982), wrote a dissertation ``Contributions to
the cosmological problem and the propagation of light in gravitational
fields'' \cite{Kohl1933} but afterwards switched to crystal symmetries
and electron theory of metals (``Kohler's rule'' for
magnetoresistance). After a professorship in Greifswald, obtained in
1943, and meteorological service during part of the war, in the 1950s
he again published on relativity theories. Max Born left 
Göttingen in summer 1933. He was known as co-founder of the new
quantum mechanics and had done only pedagogical work in general
relativity. Nevertheless, he was a good friend of Einstein and
certainly sympathetic to research in relativity theory.  Among the
philosophers, already in 1933 Hans Reichenbach in Berlin had lost his
position; likewise Ernst Cassirer in Hamburg. The climate for research
on Einstein's relativity theories became forbidding. The student from
Freiburg/Breisgau, Peter Bergmann (1915-2002), later to become an
assistant of Einstein in Princeton, in 1936 wrote his PhD-thesis with
Philipp Frank in Prague at the German University Prag on ``the
harmonic oscillators in spherical space'' \cite{MatGen1996}.\\   

Nevertheless, some isolated research went on in Germany.\footnote{This
  invalidates the opinion in \cite{Schirr2005}, where it is claimed
  that research on general relativity had completely vanished after
  1933.} In 1938, M. v. Laue reacted to a paper by H. P. Robertson on
the apparent luminosity of a receding nebula with an impact to
cosmological theory \cite{Laue1938}. In 1939, at the Physics Institute
of the Technische Hochschule Stuttgart, Helmut Hönl and A. Papapetrou came
back to the solution of Einstein's field equations for a massive
charged point particle and investigated it from the point of view of
self-energy \cite{HoePapa1939}. Papapetrou continued by looking at the
gravitational field between pole-dipole particles \cite{Papa1940}. In
1942, Otto Heckmann published his book on cosmology,  which contained a
presentation of Einstein's gravitational theory \cite{Heck1942}. Also,
a few books on the ``worldview'' of the sciences like the book by
E. Schneider \cite{Schneid1935} or the translation of Eddington's
\cite{Edding1939} and Hubble's \cite{Hubb1938}  books continued to
discuss Einstein's relativity theories and cosmology.\\  
 
But what really happened after 1933, was a definite {\em shift of
  research to the United States}. Einstein's move to Princeton was not
the main reason, although he still contributed to general relativity,
e.g., by his work with L. Infeld and B. Hoffmann on the equations of
motion of point particles \cite{EIH1938}, and with W. Pauli on the
non-existence of regular stationary exact solutions of his field
equations \cite{EinPaul1943}, \cite{Lichnero1938}. Most important was
work on gravitational collapse leading to the TOV (Tolman, Oppenheimer
Volkoff) - limit for the mass of neutron stars \cite{OppVol1939}.\\ 

The adversaries of Einstein's relativity theories in Germany,
mentioned above, now presented themselves as outright anti-Semites
battling what they named Jewish thinking.\footnote{Since 1925, Lenard,
  in an attempt to influence the German Physical Society, had already
  criticized ``the Berlin Jews'' in his correspondence
  \cite{PraesSta2012}.} According to them, quantum mechanics was also
included in such thinking. They succeeded to fill a couple of full
professorships with Nazi sympathizers, in particular Sommerfeld's
chair in Munich, and the chair for astronomy in Vienna. There, Bruno
Thüring (1905-1989) became director of the observatory and had the
time to do ``research'' on ``the Jewish question''
\cite{Thuer1941}. After he had been fired in 1945, he tried to hide
his antisemitism behind the philosophy of Hugo Dingler. Dingler in
public had attacked Hans Reichenbach as ``Einstein's favorite
philosophical associate''; cf. the article by G. Wolters
\cite{Wol1992}. In a booklet, in 1985 Thüring suggested a cosmological
model without expansion and big bang – without ever mentioning
Einstein, de Sitter or Heckmann \cite{Thuer1985}. \\      
  
\subsubsection{The war years}
After a moratorium reached in 1942 between those most fervent against
quantum mechanics and relativity theory and those who wanted to keep
physics research free of influence from Nazi organizations
(``Seefelder Religionsgespräche''), the teaching of quantum mechanics
and of general relativity no longer was obstructed \cite{Eck2007}. An
example is given by the call of C. F. v. Weizsäcker  who worked on
cosmogony to the Reichsuniversität Stra{\ss}burg in 1942
\cite{Kant1997}. Wilhelm Lenz at the University of Hamburg could give
courses on ``(Special) Relativity Theory'' both in summer 1943 and
1944 \cite{Lenz1944}. Already in 1940,  in the 2nd trimester, Hermann
Die{\ss}elhorst at  Technische Hochschule Braunschweig had lectured on
``Fundamental structures of relativity theory'' \cite{Dies1940}. A 
certain amount of publication of books dealing with Einstein's
relativity theories  continued. They mostly contained discussions of a
more general nature (natural philosophy) like in the books by Heisenberg
\cite{Heisenberg1943}, Jordan \cite{Jord1943} and Bernhard Bavink
(1879-1947) \cite{Bavi1944}, or by the theologian Arthur Neuberg
(1866-1961) \cite{Neube1941}. But there was also a textbook on
electrodynamics by Mie including relativity \cite{Mie1941}, and the
one on cosmology by Heckmann mentioned before
\cite{Heck1942}. Nevertheless, toward 1945 due to the war efforts and
destructions in Germany, serious research on general relativity had
gone into hibernation.\\        

In democratic Switzerland, courses on special and general relativity
were never considered taboo. In 1940, Andr\'e Mercier in Bern lectured
on ``General relativity theory and tensor calculus'' , and Wolfgang
Pauli on ``Relativity and gravitational theory for advanced
students''.  

\section{After World War II: Research groups and interactions}
With thirty years passed since the completion of general relativity,
Einstein's theory had to face a rival: Kaluza-Klein theory. Moreover,
Pascual Jordan had started to develop scalar-tensor
theory\footnote{Due to the destructions toward the end of World War
  II, Jordan's first paper on the subject, submitted to {\it
    Zeitschrift f\"ur Physik} {\bf 46} in 1944, had not appeared.} as 
an alternative gravitational theory. He did not know that the Swiss
mathematician Willy Scherrer (1894-1979) of Bern, in 1941 had
suggested the same theory in another context
\cite{Scher1941}.\footnote{Cf. my paper on the genesis of
  scalar-tensor theories \cite{Goenner2012}.} Einstein was fully
absorbed by his many attempts at a unified field theory. Nevertheless,  
the years following Einstein's death in 1955 brought the acceptance of
general relativity in the German speaking countries as the only
gravitational theory accepted by the majority of those involved in
gravitational research.\\   

After its defeat and division into occupied zones, Germany and Austria
were in dire straits. While universities opened rather quickly, the
Deutsche Forschungsgemeinschaft reconstituted only in 1949 and since
1951 functioned well, more or less. Einstein's Kaiser-Wilhelm-Institute
for Physics under the vice-director Max von Laue (1922-1933) and
directors Peter Debye (1935-1940) and Werner Heisenberg (1942-1945) had phased out gravitational research. In 1946, the institute reopened in Göttingen under Heisenberg and engaged in nuclear physics, quantum field theory and elementary particles. In summer of 1947, a new department for astrophysics, directed by Ludwig Biermann (1907-1986) became added. Thus, until the early 1950s, support for research on general relativity proper again could come only from single university professors.

\subsection{Federal Republic of Germany (BRD)}
\subsubsection{From 1945 to the end of the 1950s}
Since 1944, Pascual Jordan (1902-1980) had been full professor at the
University of Berlin; after 1945 he no longer could claim his position
when this university was reopened in the Soviet Sector of Berlin. He
had to wait until, after his denazification in 1947, he obtained a
guest-professorship at the university of Hamburg with the help of a
recommendation by Wolfgang Pauli. He pondered on Dirac's large
number hypothesis and on a theory of gravitation with an additional
scalar field replacing the gravitational constant in Einstein's
theory. This theory could be embedded into projective
relativity. Another one publishing in this field was Günther Ludwig
(1918-2007) at the Free University Berlin, founded in June 1948 in the
Western Sectors, and his collaborators Claus Müller \cite{LudMue1948}
and Kurt Just. From 1954 to 1956, K. Just  published eleven papers
plus one with G. Ludwig \cite{LudJu1955} on Jordan's gravitational theory.   

In 1952, in his book ``Gravitation and Cosmos'' \cite{Jord1952},
Jordan gave a summary of the results he and Heckmann's collaborators
had reached. Since 1942, Otto Heckmann had been director of the
Hamburg observatory in Bergedorf.\footnote{Cf. my entry for Otto
  Heckmann in the   Dictionary for Scientific Biography
  \cite{Goenner2008}.} He worked also on what later would be named
scalar-tensor theory and on its consequences for cosmology. Heckmann
had the advantage of being firmly established and with positions to be
filled by collaborators like the astronomer Walter Fricke (1915-1988),
later director of  the Astronomisches Rechen-Institut,
Heidelberg. Engelbert Schücking (1926-2015) who in 1952 had become a
student of Jordan, received his PhD in 1956 on non-static spherically
symmetric solutions of Jordan's vacuum field equations. He eventually
became his assistant. Together, they obtained exact solutions of Newtonian 
cosmology analogous to the Gödel metric in general relativity
\cite{HeSchue1955}, \cite{HeSchue1956}. With Schücking, Heckmann also
wrote two articles on cosmological theory (Einstein's and alternative
theories) in the new Encyclopedia of Physics (1959)
\cite{HeSchue1959a}, \cite{HeSchue1959b}. Schücking tutored Istvan
Ozsvath (1928-2013) from Hungary who obtained his PhD with Heckmann in
1960. Oszvath also published with P. Jordan \cite{JordOzs1963},
\cite{JordOzs1965}. In 1963, he left Hamburg for Texas where be became full professor in 1967 at the University of Texas at Dallas.     

In 1953-1954, after Jordan had become full professor, his seminar on
relativity and gravitation started to cooperate with Heckmann. At
the time, he also continued his long-standing interest in pure
mathematics and worked on skew lattices (special kind of algebras)
\cite{Jord1953}, \cite{JordBoe1954}. New students came in with Jürgen
Ehlers (1929-2008) becoming the most prominent, and Wolfgang Kundt
(1930- ). Their scientific contributions are already mentioned in the
improved second edition of Jordan's book which appeared in 1955  
\cite{Jord1952}. Nevertheless, in 1955 only J. Ehlers could accompany
P. Jordan to the Jubilee Conference in Bern; although O. Heckmann
lectured there on a paper with E. Schücking, the latter was not
present at the conference. G. Ludwig talked about a paper with his
coworker K. Just. In 1958, both Ehlers and Kundt obtained their
doctoral degrees with Jordan. Due to his reputation, it was not
difficult for Jordan to renew international contacts. Rainer Sachs, who 
obtained his Ph.D. at Syracuse University, N.Y. with Peter Bergmann, joined the Hamburg group as a postdoc in late 1958, just before Kundt went to Syracuse in January 1959. Kundt returned to Hamburg in the spring of 1960, when Sachs returned to the States. In 1960/1961, Dieter Brill (1933- ) from Princeton University also visited Jordan's group in
Hamburg as a post-doctoral fellow. Further guests at the group's
colloquium included well-known Wolfgang Pauli, Peter Bergmann himself, and his former student Joshua Goldberg who, from 1956 to 1963, was responsible for the United States Air Force support of research in general relativity. In 1961, E. Schücking went to P. Bergmann on a fellowship and established further contacts with astronomers and astrophysicists. The influential Alfred Schild (1921-1977), since 1957 at the University of Texas at Austin, in 1962 procured an associate professorship for E. Schücking in the mathematics department. From then on, Schücking was continuing his research outside the German relativity community. 

After the move to Munich in1958, the Max-Planck Institute for Physics with its two departments under W. Heisenberg and L. Biermann, was re-named Max-Planck Institute for Physics and Astrophysics. In the 1970s and1980s, research on general relativity (J. Ehlers) and cosmology (G. Börner) became inluded in the agenda of the department for Astrophysics.

\subsubsection{From the 1960s to the 1980s}
Another early member of the group since 1957 was Manfred
Trümper\footnote{He is a younger brother of the well-known
  astrophysicist Joachim E. Trümper.} who joined the seminar in 1957
and obtained his PhD with Jordan in 1962 \cite{True1962},
\cite{TrueKu1962}. Already in 1958, J. Ehlers and K. Kundt showed, by
their doctoral theses, that the emphasis in the group's work had
shifted to mathematical aspects of general relativity, including the
search for exact solutions.\footnote{J. Ehlers wrote his dissertation
  on ``Construction and Characterization of the Einstein equations.''}
Coordinate-free covariant methods were developed and applied to the
definition of radiation and the description of fluid matter. The
results were published from 1960 to 1965 in the proceedings of the
Academy of Sciences and Literature in Mainz.\footnote{Some of them
  were translated into English and reprinted as ``Golden Oldies'' in
  the journal of General Relativity and Gravitation
  \cite{JorEhKu1960}\cite{JoEhSa1960}, \cite{JorEhSa1961},
  \cite{JorKu1961}.} Jordan still worked out consequences from Dirac's
large numbers hypothesis for geophysics: the Earth should have
expanded during its past \cite{Jordan1966}; in addition to the
previous acknowledgments, now Manfred Trümper is also in the list of
contributors. He had become post-doc with Peter G. Bergmann at
Syracuse University in the academic year 1962/63, and in 1963/64 at
Yeshiva University in New York. In the years 1965 to 1969, he was
assistant to P. Jordan.\\  

Thus, the ties between Peter G. Bergmann  and Pascual Jordan with
regard to the exchange of post-doctoral fellows were already tight
when Jordan visited the US (New York) in 1963. Due to these frequent
exchanges between some of the members the Hamburg group and young
colleagues of the United States, Jordan was forced to raise a
considerable amount of money. A big spender was Friedrich Flick
(1883-1972), one of the richest men in the BRD with an ugly Nazi
past. Other sources were the Volkswagen Foundation, the Fritz ter Meer
Foundation, or the Academy of Science in Mainz. The contracts with the
European Office of the Aerospace Research USAF between 1958 an 1967
were very important. In 1964, Jordan and four of his young
collaborators wrote a (third) final 
report of more than 100 pages ``Contributions to Actual Problems of
General Relativity'' for this monetary source. He also needed some
time  for his role as a member of Parliament in the Bundestag in Bonn
from 1957 to 1961. One may safely assume that most of the research
work in gravitation was actually achieved by his younger collaborators
who welcomed  his name for making themselves known. At least until the
end of the 1960s, the Hamburg group was the largest 
and most influential of the groups in Germany working on general
relativity. In 1966, P. Jordan proudly stated about his Hamburg
seminar: \begin{quote} ``The development of this seminar was able to
  contribute to the accomplishment of the task of keeping alive the
  German participation in research on relativistic gravitational
  theory. There was a danger in this respect, because in the preceding
  period relativity theory had stood back greatly in the awareness of
  German physicists. When the new beginning started, nuclear physics
  and elementary particles were outshining relativity.''\footnote{Die
    Entwicklung dieses Seminars konnte einen Beitrag leisten zur
    Lösung der Aufgabe, in der Bundesrepublik die deutsche Mitarbeit
    an der relativistischen Schwerkraftforschung nicht einschlafen zu
    lassen. Gefahr in dieser Richtung war gegeben, da in der
    vorangegangenen Zeit die Relativitätstheorie im Bewusstsein der
    deutschen Physiker stark zurückgetreten und beim Neuanfang
    zunächst von Kernphysik und Elementarteilchen gewissermaßen
    überstrahlt war.} \end{quote}

\subsubsection{The first step toward institutionalization of gravitation research}
A step toward the institutionalization of gravitational research in
West-Germany was made only at the beginning of the 1970s, roughly
simultaneously with the foundation of the International Society on
General Relativity and Gravitation in Copenhagen in 1971. The Hamburg
Seminar faded after P. Jordan had retired and the Hamburg faculty did
not replace him by a colleague who would continue research in
gravitation. With Lehmann and Döring on the faculty, research and
teaching in the field of gravitation gravitation was deemed
unnecessary. \footnote{As is clear from his correspondence, Jordan had
  favored Jürgen Ehlers to become his successor \cite{Jordan1967} I owe the
    knowledge of this letter to Dr. A. Blum, Berlin.}. Whether the
  disdain of the faculty for the topic of gravitation, or unease about
  Ehler's monopolizing personality were decisive, remains open. As one
  of the  most active members of the group, J. Ehlers, from 1964 until
  1971 mostly worked in the United States, first from 1964 to 1965 at
  the   University of Texas in Dallas and then from 1966-1971 at the
  University of Texas in Austin.\\   

Also W. Kundt had been attracted to the US: as a Flick exchange fellow
he went to Syracuse in January 1959 where Rainer Sachs was to receive
his PhD; Sachs then joined the Hamburg group for a year. Kundt
returned to Hamburg in the spring of 1960, when Sachs returned to the
United States. Since 1971, he had become titular professor in
Hamburg. Jordan's PhD-student H.-J. Seifert was tutored by him
(1969).\footnote{Seifert eventually became professor at the Hochschule
  der Bundeswehr (now Helmut Schmidt University) in Hamburg like
  another doctoral student of Jordan, Henning Müller zum Hagen.} Yet, 
from 1972 on he changed the field of research and went into
astrophysics in Bonn. Nevertheless, until 1975 he tutored students in
Hamburg, applied and received support for them by Deutsche
Forschungsgemeinschaft. From 1968 to 1969, M. Trümper was drawn to the
University of North Texas in Denton, north of Dallas, and from 1969 to
1974 to Texas A\&M University in CollegeStation. After some time at
the Max-Planck-Institute for Astrophysics in Ehler's department (see below), he worked in several countries on three continents. At the time, after World War II, even a little known place in academia, in particular for gravitational theory, like the
University of Texas at Austin or Dallas, held advantages for Jewish
emigrants like Wolfgang Rindler and Alfred Schild (1921-1977), and
for West-German relativists like those of the Hamburg group. They
achieved a lot; e.g. in 1963, in order to make the University of
Dallas known in the scientific world, the ``Texas Symposium [later
``Conference''] on Relativistic Astrophysics'' was founded by Alfred
Schild, Engelbert Schücking (1926-2015) and Ivor Robinson (1924-2014) \cite{Schueck1989} which celebrated its 50th anniversary in 2013 \cite{Salis2013}.\\   

In the group around G. Ludwig in Berlin, a PhD thesis of 1962 by Karl
Kraus (1938-1988) written under the guidance of Kurt Just investigated
a Lorentz-invariant gravitational theory. In 1961 K. Just left for the 
University of Arizona in Tucson and the work on gravitation finally
came to an end around 1963 when Ludwig took a position at the
University of Marburg and Kraus went with him as his assistant. Other
smaller groups developed at the Technische Hochschule Braunschweig,
around Max Kohler and at the University of Freiburg/Brsg. around
Helmut Hönl (1903 -1981) and his assistant Konradin Westpfahl
(1926-1994), at the time a lecturer. As their publications show, for
both, Kohler and Hönl, at first gravitation formed a side issue
compared with transport theory and diffraction optics \cite{Koh1949},
\cite{HoeMaWe1961}, \cite{West1955}. Between 1951 and 1953, Kohler
published four papers on general relativity and bi-metric theory. Hönl
had a paper on Mach’s principle in 1953 and in 1955 a single paper on
the gravitational field of rotating masses with A. W. Maue
\cite{HoeMau1956}. After his call to the university of Göttingen,
Kohler continued to work in both fields while Hönl in Freiburg turned
mainly to general relativity, studying, among other themes, with
H. Dehnen the role of Mach's principle, and with K. Westpfahl
equations of motion of point particles. In 1970, H. Dehnen became the
only one appointed as full professor in the field of relativistic
gravitational physics in the Federal Republic (BRD) after Jordan and
until today (University of Konstanz). Since the mid 1960s, Friedrich
W. Hehl, Technische Universität Clausthal, introduced research on
spin-angular momentum within theories of gravitation with torsion and,
after his appointment at the university of Cologne in 1975, built up
an active research group directed, among others, to Poincar\'e 
gauge theories of gravitation. In Würzburg, the theoretical physicist
R. Ebert (1926-2013) guided a group working in relativistic
astrophysics from which  R. Breuer, W. Dietz and, through his
habilitation, E. Hilf emerged.\\   

The situation of gravitational research in BRD (and to lesser extent
in GDR) reflected itself, more or less properly, at the meeting of the
German Physical Society from Oct. 4-9, 1965 in Frankfurt am Main
(Jahrhunderthalle der Farbwerke Höchst AG). Included were a plenary
talk ``Neuere Entwicklungen in der Allgemeinen Relativitätstheorie''
by J. Ehlers (Dallas), on Friday, Oct. 8, a topical session
(``Fachsitzung B'') under the chairmanship of Pascual Jordan, and an 
additional meeting during which talks were presented which, in the
program, were listed under the heading ``Further contributions not to
be presented orally''. The topical session contained two Fachberichte
by G. Ludwig, Marburg and H. Hönl, Freiburg as well as 8 further short
talks by people from Hamburg (3), Freiburg (3) and from GDR
(Schmutzer, Treder). The five additional presentations came from
Hamburg (3) and Freiburg (2) (\cite{DPG1965}, p. 61, 64-67). I imagine
that the initiative started from P. Jordan who invited other groups to
join. That only 2 representatives from GDR were there, the leaders of
the 2 main groups in Potsdam and Jena, possibly was due to the very
restrictive policy concerning traveling into countries outside the
``iron curtain''. I visited the meeting and took along my adviser
Prof. K. Westpfahl with the car lent from my parents.\\ 

In 1965, the 50th anniversary of the completion of General Relativity
was celebrated in Berlin both in East and West, but now
separately.\footnote{The Berlin Wall finally separating both
Germanies had been erected in August 1961.} In East-Berlin, H.-J. Treder and the Academy - whose name in 1972 would be changed into Academy of Sciences of GDR - correspondingly organized a big international symposium \cite{Tred1966}.\\

Following intensive debates during the Frankfurt-meeting of the German
Physical Society, a public dispute between the assertive J. Ehlers
and E. Schücking on the one side, and H. Dehnen on the other, surfaced
concerning both the physical interpretation of an exact solution of
Einstein's equations, and Mach's Principle \cite{Ehschu1967},
\cite{HoeDe1967}. As a consequence, the then co-editor of Zeitschrift
für Physik, Nobel prize winner H. Jensen, decided to stop printing
articles on general relativity.\footnote{Interestingly, in the 1950s,
  S. Goudsmit then editor of Physical Review also intended to ban
  papers on general relativity from this
  journal. Cf. \cite{DeWitt2009}, p, 414. This reference has been
  taken from \cite{Kragh2016}.}  It is possible that this damaged  
Ehler's image at universities in BRD, because his attempts to obtain a
full professorship at a university seemingly failed. In 1971, through an initiative of the
astrophysicist Ludwig Biermann (1907-1986), director at the
Max-Planck-Institute for Physics and Astrophysics in Munich, Jürgen
Ehlers was invited to join this institute as director of a permanent
working group on gravitational theory. The importance of Biermann’s initiative may be seen in that it brought astrophysicists and relativists into closer cooperation. At the same time, Ehlers  was named
honorary professor at Ludwig-Maximilian University. His group became
the first permanent anchorage ground for gravitational research in
West-Germany independent of university financing. Since fall 1979
until 1982, M. Trümper joined the group in Garching. Other members of
the Hamburg group like B.G. Schmidt who had written his dissertation
with P. Jordan and developed the concept of b-boundary of a manifold,
or H. Friedrich, joined Ehlers on permanent positions.\\       
While the establishment of Ehler's group definitely represented progress for the standing of the field of gravitation within the German physics community, it also aroused some jealousy among the small groups at the universities. This was due to the much better financial means provided by the Max-Planck-Society for the organization of meetings, invitation of guests from abroad, travel to conferences etc. Above all, the positions at Max-Planck-Institutes were full-time research  positions with no teaching obligations.\\ 

At the joint meeting of the German and Austrian Physical Societies
and the German Geophysical Society in Salzburg, 29. Sept. to
4. Oct. 1969, the ``Festvortrag'' was held by W. Thirring on 
``Gravitation''. A topical meeting ``Relativity theory and cosmology''
took place under the joint chairmanship of Max Kohler, Göttingen, and
Roman Sexl, Vienna. There were talks by  N. (?) Bondi, (Neuilly-sur
Seine) [Must really have been Hermann Bondi, cf. his ``Gravitational
bounce in general relativity'', {\it Monthly Notices of the Royal
  Astronomical Society} 02/1969; 142], and F. Pirani, as well as 2
further talks from the Hamburg group (H.-J. Seifert, O. Störmer) and a talk by J. K. Lawrence from the theoretical physics
institute of the University of Vienna. As an assistant of M. Kohler, I
attended the meeting (\cite{DPG1969}, p. 683, 688-690).\\  

Thus, until the 1970s, the situation in the Federal Republic of Germany
was very much the same as the one before the war: around some single
professors small groups were pursuing gravitational research. This
was not exactly a stable situation; financial support came from
universities and personal applications to the German Science
Foundation. In 1973, during an international Symposium in Bonn,
P. Jordan pointed out that ``[..] the present state is such that [..]
in the area of the Federal Republic the theory of General Relativity
does not at all receive the deserved recognition and research on it is
 not adequately continued'' ({\cite{Bleu1973}, p. 2). As this
   symposium and a preceding one in 1971 shows, there was some general
   interest also among mathematical physicists.\\   
In 1978, an international conference, the 9th Texas Symposium on
Relativistic Astrophysics, was held in Munich under the auspices of
the Max-Planck-Institute for Astrophysics \cite{EhPeWa1980}.   
On 2 March 1979, under the auspices of the German Physical Society, a Gedenkveranstaltung zum 100. Geburtstag von Albert Einstein, Max von Laue, Otto Hahn und Lise Meitner'' was held in Berlin with J. A. Wheeler, Austin presenting a talk ``Einstein und was er wollte''
\cite{DPG1979}, \cite{Wheel1979}. 

Inspite of all the celebrations, until the late 1970s only about 20
permanent positions for scientists doing research in general
relativity were available in the Federal Republic of Germany. Among
them, three full professorships (Freiburg, Konstanz, Würzburg) and
three associate professorships (Göttingen, Köln, Bonn). We may add the
position of J. Ehlers at the Max-Planck-Institute for Physics and
Astrophysics in Munich and a professorship at the Bundeswehrhochschule
in Hamburg \cite{Breuer1979}.     

\subsection{German Democratic Republic (GDR)}

\subsubsection{The first decades until 1980}
At its beginning, in the German Democratic Republic a very different
situation obtained. The predominance of the ``working class'', as
forcefully set by the party in power, SED (Sozialistische
Einheitspartei Deutschlands), resulted in some hostility against
academic labor  which until then was anchored in the middle class
(bourgeois intelligentsia). But for an efficient economy, the old
and many new university graduates were needed: An aspired goal of the late 60s was to educate until the 1990s the majority of the
workforce at advanced technical colleges and universities
(\cite{Meyer1990}, p. 7).  In GDR, two ministries were responsible for research and teaching, the ``Hochschul''-, and ``Wissenschaftsministerium'', at times rivaling each other.\\

In this context, the German Academy of Sciences in Berlin was given a
leading role for research, particularly in the exact sciences, following the example of the Soviet Academy of Sciences. With Albert Einstein as one of its
former prominent members, the party’s intention was to continue
research on his theory. With the appointment of Achille Papapetrou, in
1952, to the Research Institute for Mathematics of the German Academy
of Science, a seed for research and teaching in general relativity was
planted. Papapetrou, who due to his leftist political opinions had
been dismissed from his professorship in Greece, came from a temporary
position with Leon Rosenfeld in Manchester. He had been recommended by
Einstein's former collaborator E. Freundlich\footnote{He had changed
  his name to Finlay-Freundlich by putting his mother's name in
  front.} who had emigrated to England and who had been consulted
during the preparation of a solar-eclipse-expedition by the Berlin
Academy. Freundlich did not want to become totally involved and
suggested A. Papapetrou as a coworker. Papapetrou’s position in
Manchester ran out at about the same time \cite{Hoff2016}. In August
1951, he obtained a position in the Research Institute for Mathematics
of the German Academy of Sciences in Berlin and in 1953 became head of
a research group for mathematical physics \cite{Papa1955}. In 1957,
Papapetrou was promoted to professor at Humboldt University in
Berlin. One of his first doctoral students was Hans-Jürgen Treder
(1928-2006); he obtained his PhD in 1956, his habilitation in 1960 on
shock waves and became a heavyweight in gravitational research within
the Academy. He had been a member of the communist party in
West-Berlin; his contacts with influential members of the State's
party like Kurt Hager (1912-1998), member of the Central Committee
responsible for all cultural affairs, and the well known physicist
Prof. Robert Rompe (1905-1993)\footnote{Treder is coauthor of 7 books
  with Rompe.} made 
him politically unassailable. After the construction of the ``Berlin
Wall'', Treder left West-Berlin to settle permanently in East-Berlin
or, as it was then called by the GDR-authorities: ``the Capital City
of GDR''. Another doctoral student of Papapetrou was Georg Dautcourt
at the Institute for Pure Mathematics of the Academy
\cite{DaPaTre1962} who at first could establish a research group but
which in 1971 was dissolved after he had criticised Treder's research
agenda. Eckhard Kreisel also belonged to Papapetrou's doctoral
students but could not finish the work before Papapetrou left.\\  

Papapetrou left GDR in 1961/62 for Paris. In the aftermath,
H.-J. Treder  became professor for theoretical physics at Humboldt
University in Berlin and from 1963 to 1966 director at the Institute
for Pure Mathematics of the Academy. As a consquence of the 3rd
University- and Academy-Reform of 1968 in GDR \cite{Wolter1993},
several astronomical observatories and astrophysical institutes became
merged under the umbrella of the Academy. Since 1969, Treder was in charge of the newly launched ``Central Institute for Astrophysics''
(ZIAP). The ``Central Institute'' belonged to the research domain
``Cosmic physics'' of the Academy and was focused more on
experimentation/observation. With his function, Treder obtained a seat
in the steering committee of the Academy. He was very productive until
his death; some of his hundredths of publications are listed in an
obituary \cite{Schim2008}. His research was directed toward
unification of gravitation and electrodynamics, alternative
gravitational theories (tetrad theory, curvature squared Lagrangians),
shock waves in Einstein's theory, and a mechanics without inertial
mass. As Treder's research topics kept aloof from the mainstream of
international research in gravitation and mainly published in
German, most of his papers and books had a very limited influence on
international developments in general relativity.   

Earlier, gravitational research had been followed also at other
universities of GDR, e.g., in Greifswald and Leipzig by Dietmar
Gei{\ss}ler, Hans-Georg Schöpf and Adolf Kühnel, respectively. Schöpf
was transfered to mathematical physics in Dresden; Kühnel went into
condensed matter physics. Also, in Leipzig around P. Günther in the
mathematics department, interest on particular problems related to
Einstein’s equations originated. His doctoral students R. Schimming,
then professor in Greifswald and V. Wünsch (1941-2015), professor at
the Pedagogical University Erfurt, and after retirement in the faculty of
mathematics of the University in Jena, took up this work. In the 1970s
and 1980s, we can speak of an institutionalization of research on 
relativity and gravitation in terms of a stable structure with two
centers. That the two groups were not really cooperative but rather in
competition is another story.\footnote{The situation may be mirroring
  the relationship between Wissenschafts- and Hochschulministerium of GDR with the first one being closer to Treder.} An idea envisaged
with the 3rd University- and Academy-Reform was that research and
teaching should be separated – with the institutes of the Academy
responsible for research and the universities for teaching. The
example of the University Jena shows that this idea could not be
enforced in practice. At the beginning of the 1970s, the original
demand to bind research in the natural sciences at the universities to
industrial research became also relaxed. (\cite{Schramm2008}, p. 147)
Nevertheless, the Institutes of the Academy of Science of GDR had
much better resources; e.g., in general twice as many positions than the universities \cite{Daeum1993}.\\ 

After having obtained in 1955 his PhD with Hans Falkenhagen at the
Wilhelm-Pieck-University Rostock, Ernst Schmutzer became his assistant
and in 1956/57 gave the first courses on general relativity in Rostock
after World War II. In 1957, he moved to the Friedrich-Schiller-University
of Jena, obtained his habilitation there in 1958, and was appointed
professor in 1960. With a couple of master-degree students, he
managed, during the 1960s, to establish a group for research in
general relativity and 5-dimensional gravity \cite{Schmutzer1968},
\cite{Schmutzer2015}. His request to the ministry in Berlin, i.e., to
approve a main focus ``relativity and gravitation'' in Jena, was
accepted. Members of the faculty\footnote{Mathematisch-Naturwissenschatliche Fakultät; between 1968 and 1990 ``Sektion Physik''.} working in general relativity
and/or in astrophysics were Dr. Hans Stephani (1935-2003);
Dr. Nicolaus Sali\'e; Dr. Dietrich Kramer, Dr. Gernot
Neugebauer\footnote{In fact, Neugebauer was working with Gerhard Kluge
  on relativistic thermodynamics and not directly related to
  Schmutzer.}; Dr. Eduard Herlt; Dr. Rainer Collier; Dr. K.-H. Lotze
among others.\footnote{No   distinction has been made between doctoral 
  titles like Dr.rer.nat.habil. or Dr. sc.nat. granted in GDR during different
  periods. Also, differing positions, like lecturer or assistant
  professor are not shown.} 
Important results were obtained, in particular with regard to exact
solutions of Einstein's field equations, by G. Neugebauer,
H. Stephani, E. Herlt and D. Kramer, among others. E. Schmutzer and 
his colleagues, within the framework of the Relativity Seminar at the
University of Jena, also organized international colloquia in
Georgenthal, Thuringia, where scientists from both sides of the ``Iron
Curtain'' could meet and discuss, e.g. the 14th Seminar, 15. -
21. Nov. 1882,  or the  15th Seminar 26. 11. -
2. 12. 1984.\footnote{Only a few persons from this group, belonging to
  the so-called travel-cadre (Reisekader) were permitted to travel
  into Western countries.} Before the end of the 1980s, Reinhard Meinel joined the group on general relativity in Jena .\\

\subsubsection{The last decade before the end of GDR}
In 1979, for the 100th  anniversary of Einstein's birthday, conferences were organized in both parts of Berlin. In West-Berlin the title of the
conference was ``Einstein Symposium'' and its contributors were many
of the leading relativists of the ``West'', politically speaking
\cite{Nel1979}. Some of them like J. A. Wheeler gave a lecture both in
East and West.

One year later, in 1980, the only large {\em international} conference in
Germany under the auspices of the ``International Society of
Relativity and Gravitation'', GR 9, was organized by E. Schmutzer and his co-workers and colleagues in Jena in GDR, not by Treder's
group. According to those involved in Jena, it took quite some
negotiations with the Party and the Ministry before GR 9 could be run in Jena.\footnote{In 2016, GR 21 will be held in New York City.} \\

Following the ``Forschungs-Verordnung'' of June 1986, the universities in GDR were again obliged to use a minimum of 50\% of their research-potential for industrial research and development
(\cite{Meyer1990}, p. 11). This was due to the precarious financial
situation in GDR. In 1982, Treder handed over the ZIAP to his
successor, the astronomer  Karl-Heinz Schmidt (1932-2005), due to
health problems. He was appointed director of a newly founded
Einstein-Laboratory for Theoretical Physics in
Potsdam-Caputh\footnote{With two sites at the observatory in
  Babelsberg and in the former summer house of Albert Einstein in
  Caputh.} and remained in this position until 1992. Treder's
Institute included: Dr. Dierck-Ekkehard Liebscher,
Dr. Eckhard Kreisel who likewise had written his thesis in 1965
under Treder's tutoring\footnote{Both were titular professors.},
Dr. Horst-Heino von Borzeszkowski who had received his PhD with Treder
in 1973 and Dr. habil. Ulrich Bleyer. Dr. R. W. John from ZIAP,
although associated with Treder, scientifically went his own way.  
 
A philosopher of science, Dr. Renate Wahsner, for some time also
belonged to the Einstein Laboratory. Further coworkers in general
relativity were Dr. H. Fuchs, Dr. S. Gottlöber, Dr. U. Kasper,
Dr. J. Mücket, Dr. V. Müller and Dr. habil. H.-J. Schmidt. 
 Until 1986 when he left GDR, also Prof. Dr. Helmut Günther (1940- ) had worked in both the Zentralinstitut für Astrophysik and the Einstein-Laboratory for Theoretical Physics\footnote{In 1972, he had obtained his Dr. sc. nat.; he later published on disorder in lattice structure  and on Lorentz-symmetry.} 

Thus around 1979, in GDR manpower in terms of salaried positions
surpassed the one in FRG; this includes leading positions. In Treder's
group (Zentralinstitut für Astrophysik and Einstein-Laboratorium) one
full professorship existed (H.-J. Treder, member of Academy of
Science) plus two senior collaborators D.-E. Liebscher, E. Kreisel. At
the University of Jena E. Schmutzer was full professor and group
leader with two further senior collaborators, D. Kramer, H. Stephani
and another one in relativistic thermodynamics (G. Neugebauer). In
FRG, the field contained only one full professorship (H. Dehnen,
Konstanz) and three associate professors (H. Goenner, Göttingen; F. W. Hehl, Köln; K. Westpfahl, Freiburg).

\subsection{Foundation of a subdivision within the German Physical
  Society in BRD and its impact.}  

A first sign toward formation of a ``community'' showed up in June
1971 during a ``Colloquium on Relativistic Astrophysics'' in the
observatory ``Hoher List'' under the patronage of German Science
Foundation (DFG). New astrophysical objects like quasars, pulsars and
neutron stars had came into the center of
interest. M. Reinhardt\footnote{M. Reinhardt later became professor at 
  the University of Bochum, but died young in 1985.} of Bonn
University gave a report of the 5th Texas Symposium on Relativistic
Astrophysics at the University of Texas at Austin in Dec. 1970. As a
consequence of the discussions, a request for the establishment of a
priority program ``Relativistic Astrophysics'' to be funded by DFG was
drafted. It became implemented, supervised by Erich Kirste (1927-2002)
from DFG, and coordinated by J. Ehlers, Munich during the 5-year
period of 1974-1979. Of the total of 71 funded projects, 57\% (41) went to
astrophysics while only 43\% (30) were relevant to general
relativity. Half of the support for these 30 research projects went to
the two biggest groups in Hamburg (W. Kundt, B. G. Schmidt) and in
Munich (J. Ehlers) -- both offspring of Jordan's
seminar. It is true that in particular some scientists with a
background in nuclear physics working for the understanding of neutron
stars, like Peter Mittelstaedt (1929-2014), Cologne, and Konrad Bleuler
(1912-1992) in Bonn, took also advantage of the priority
program. Nevertheless, one of the intentions of this program, i.e., to
awaken the interest of more universities toward establishing research
groups or positions in (relativistic) gravitation turned out to be a
failure \cite{Breuer1979}.\\                                    

The initiative for the foundation of a subdivision (``Fachverband'')
``Gravitation and Relativity'' within the German Physical Society (DPG)
arose from informal discussions among J. Audretsch (Konstanz), H. Goenner (Göttingen) and F. W. Hehl (Köln). After H. Dehnen (Konstanz) had been convinced of the idea and had contacted J. Ehlers, a meeting on general relativity was organized in September 1983 by J. Ehlers, B. G. Schmidt and M. Walker, supported by
the Max-Planck-Institut for Physics in Munich. It was held at Castle
Ringberg near Tegernsee (\cite{JahrberMue1983}, p. 223). Now, the wish to organize gravitational research within the German Physical
Society was communicated and backed by the participants.  Further
discussions took place on 22./23 June 1984 during a
Köln-Göttingen-Colloquium  in Cologne. The actual foundation of the
Fachverband took place on October 3/4, 1984 in the conference center ``Physikzentrum Bad Honnef'', in the presence of the then president of DPG, Prof. J. Treusch, after the DPG had become convinced that a sufficient
number of ``relativists'' would become members. As H. Dehnen, at the
time dean in Konstanz, did not run for office and F. Hehl was away in
the USA, J. Ehlers became elected first president. He organized a
first working meeting on ``Gravitation and Relativity'' in Bad Honnef
from 8.-13. Dec. 1985 with a sizeable number of participants, also
from Switzerland and Austria.\footnote{At the University of Vienna, Roman U. Sexl (1939-1986) was professor for Cosmology and General Relativity since 1972. The joint work with his colleagues in Vienna, Peter Aichelburg, on the ultraboost of the Schwarzschild vacuum \cite{AiSex1971} and Helmuth Urbantke on cosmic particle creation \cite{SexUr1967} as well as his many books are well-known \cite{SexUr1992}, \cite{SexUr2002}. Two further, more mathematically
inclined, colleagues of the group must be presented, i.e., Robert Beig
and Piotr T. Chrusciel. Unfortunately, the intended section concerning research on general relativity in Austria and Switzerland still waits to be written.} Further such meetings followed every two years, partly as WE-Heraeus-Seminars. The chairmanship of
J. Ehlers who had been re-elected once, lasted until 1989, when Gerhard Schäfer, Munich, succeed him during the 3rd working meeting of the Fachausschuss \cite{Goenner1990}.\footnote{Gerhard Schäfer had been a student of H. Dehnen in 1973-1986 and post-doc with J. Ehlers in 1988-1992.} After G. Schäfer, F. Hehl, Cologne,  became the next chairman. Thus, also in
the Federal Republic of Germany (BRD), a stable reference base for
research on general relativity had come into being. Apart from the
organization of its own meetings, the main activity of the
Fachausschuss (later renamed Fachverband) within DPG consisted in
taking part in the yearly meetings of the German Physical Society by
supplementing plenary talks and subsections of the program. This was
organized by the chairmen at the time. Relativists thus had a
continuous platform for presenting their work. It should be noted that there is no corresponding subdivision for astrophysics within the DPG; this field finds its base in the German Astronomical Society.\footnote{The title of this year’s (2016) 89th general assembly  shows this clearly: ``The many Facets of Astrophysics - Photons, Particles, and Spacetime.''}
  
\subsection{Situation after German re-unification 1990}

After the German Democratic Republic, on October 3,
1990, joint the Federal Republic of Germany, on November 20, 1990, the Physical Society of the GDR also merged with the DPG. A restructuring of the scientific landscape of the former GDR was started, in particular of
the Institutes of the Academy of Science. They were given the
guarantee that operation would be financed until Dec 31. 1991. It was
unclear what would happen to the Zentralinstitut für Astrophysik and,
particularly, to the Einstein-Laboratory for Theoretical Physics, in
both of which research on general relativity had been done. In all
likelyhood, the Einstein-Laboratory would not be continued despite
the pleading of international supporters like Peter Bergmann of
Syracuse: H.-J. Treder had become permanently ill, passed his zenith,
and -- in the eyes of West-German opinion leaders -- was compromised by
his closeness to the ruling party SED. In retrospect, it seems clear
that, by this connection, he had been able to play an important role in
securing financial support for research in gravitation.\\   
A phaseout of Treder's group in Potsdam followed, with some of its
members like U. Kasper, H.-H. v. Borzeskowski and H.-J. Schmidt then
receiving 2-year-contracts for pursuing research in gravitation under
a program with the typically German name
``Wissenschaftler-Integrationsprogramm'' (integration program for
scientists). WIP ran out at the end of 1996. Eventually, some of 
those working in the Zentralinstitut für Astrophysik, like H. Fuchs,
S. Gottlöber, D.-E. Liebscher, J. Mücket and V. Müller, could join the
Leibniz-Institut für Astrophysik in Potsdam. Dr. Bleyer became head of
Urania in Berlin.\footnote{Urania, founded in 1888, is a privately
  organized association the aim of which is to communicate the most
  recent scientific findings to the broad public.} H.-J. Schmidt
joined the mathematics department of the newly established University of
Potsdam.\footnote{From 1996 to 2005, he was editor of the scientific journal 
  {\it General Relativity and Gravitation}, published since 1970 under
  the auspices of the International Society on General Relativity and
  Gravitation.}   

In contrast, gravitational research in Jena could adapt more softly to
the new situation. E. Schmutzer's co-workers with salaried positions
could stay in the physics department of the university (Stephani,
Kramer, Salié, Herlt, Collier). G. Neugebauer was invited by the
president of the Max-Planck-Society to establish in Jena one of the 28
working groups supported by the Society for a duration of five
years. The members of the group chosen by Neugebauer were R. Meinel
and A. Kleinwächter, both of Jena, W. Kley, at the time in the United
States, and Gerhard Schäfer from Ehler's group in Garching/Munich. In
1996, the group became absorbed by the physics department of the
university.\\      
A reorganization also took place in former West-Germany: In 1991, the Max-Planck-Institute (MPI) for Physics and Astrophysics was split up into three independent institutions, the MPI for Physics, the MPI for Astrophysics and the MPI for extraterrestrial Physics, originally established as a subdivision in 1963. The working group ``Gravitational theory'' under its director J. Ehlers continued in the MPI for Astrophysics until 1995.\\

Right after the German`` re-unification'', on February 8, 1991,
Friedrich Hehl, Cologne and Hubert Goenner, Göttingen, formulated a
Memorandum on the foundation of an International Einstein Center in
Potsdam/Caputh. 
 \begin{quote} ``Quite certainly, no 'relativist' will be appointed
   to the full professorships mentioned above after retirement of the
   present incumbents. As seen from the international standard of
   competition in a fundamental branch of modern physics, for junior
   reseachers this situation is, consequently, rather discouraging in
   terms of job openings etc. A closing down of the
   Einstein-Laboratory, without substitution, would appear
   irresponsible under such circumstances.''\end{quote} 

This memorandum was submitted to the secretary of the German Council of
Science and Humanities (Wissenschaftsrat) on Feb. 1991 and later
announced in the journal ``Physikalische Blätter'' related to the
German Physical Society \cite{GoeHe1991}\\ 
The original plan to establish a joint German-Israeli
research institute won the approval of the famous Israeli theoretical physicist
Yuval Ne’eman (1925-2006);\footnote{Y. Ne’eman co-discovered
  SU(3)-symmetry in particle physics. He was president of Tel Aviv
  University (1971-1975) and founder of the Israel Space Agency in
  1983.} the intention was to approach the German-Israeli Foundation for
Scientific Research (GIF) to take part in this initiative besides the
German Federal Ministry for Science \& Technology and the State of
Brandenburg. It turned out, however that, due to its bylaws, GIF could
not permanently support such an Einstein Center. Also, the State of
Israel was short of money to be used for an eventual establishment of
the necessary two Einstein Centers, one in Jerusalem, the other in
Germany. \\      
Fully aware of this initiative and of its missing institutional
background, in July 1991, the German Council of Science and Humanities
suggested to the Max-Planck Society the establishment of a working group for the
eventual foundation of an Albert-Einstein-Institute on a national
basis. The working group was then formed with Jürgen Ehlers as its
chairman. This way of proceeding was supported by the Fachverband
Gravitation and Relativity through its chairman, G. Schäfer, and by
the then president of DPG, Th. Mayer-Kuckuk (1927-2014). To the
physics community, neither the members of this group nor its
proceedings were communicated. It was not before the summer of 1993
that a ``Memorandum on the founding of an Albert Einstein Institute
für Gravitationsphysik'' was issued and invitations for comments from the community during a ``Symposium on Developments and Trends in
Gravitational Physics'' held on Sept. 20-21, 1993 in Munich sent
around. Max-Planck Society had introduced a ``Scientific Organization
Committee'' with J. Ehlers and the directors of two other
Max-Planck-Institutes among further individual members. The
memorandum emphasized:\begin{quote} ``What is missing is an institute
  where researchers from Germany and abroad can collaborate for
  reasonable periods of time. An Einstein Institute could serve this 
  purpose and thus stimulate also both research and teaching at
  universities. Universities cannot play this role: Positions are not
  available, high-level teaching requires a minimal number of people
  with small teaching obligations working in close contact with each
  other and with guests from abroad.''\end{quote} Thus, similar to
what had been formulated during the 3rd Hochschulreform of GDR in
1968, the intention was to clearly distribute tasks between low-level
teaching at universities and research-oriented high-level teaching in
close cooperation with Max-Planck-Institutes. The founding of the
Institute also accepted the lack of positions for relativity research
in Germany as unalterable. As it turned out, only those junior
scientist already inside the Max-Planck-Society would obtain permanent positions in the new institute. \\    

Eventually, the new Max-Planck Institute for Gravitational
Physics \linebreak (``Albert-Einstein Institute'') opened in Potsdam in
1995. J. Ehlers, Munich and B. Schutz, Cardiff became two of the
directors of the newly founded Institute for Gravitational Physics 
with three departments.\footnote{Their third colleague H. Nicolai,
Hamburg, joined them in 1997.} Unfortunately, by its construction
the structure of the Albert-Einstein Institute showed that general
relativity was considered as an appendix to either astrophysics and
elementary particle physics, or to mathematics. With countless guests
from abroad, the institute established a leading international role in
research. Yet, the job-situation in Germany was not improved by the
Albert Einstein Institute. The steady number of about 14 PhDs produced per decade in the field of general relativity, cosmology and
relativistic astrophysics during the 1960s to the 1980s, remained
uninfluenced by the new Max-Planck Institute for Gravitational Physics. 
In 2002, to the Albert-Einstein-Institute a fourth section on
experimental gravitation (measurement of gravitational waves by
interferometry) in Hannover (K. Danzmann) has been added -- with
recent great success. 
\section{Conclusion}
At present, an institutionalization of research in relativistic gravitation has been achieved in Germany through a thriving topical section
``Relativity and Gravitation'' of the German Physical Society and a
Max-Planck-Institute for Gravitational Physics. This topical section
of DPG belongs to the smaller ones in comparison with elementary
particle physics, physics of surfaces, semiconductor physics, nuclear
physics, or quantum optics, etc. \footnote{A union of 6 sub-divisions
  of the DPG including the Fachverband ``Relativity and Gravitation''
  with the name ``Matter and Cosmos'' has been established in 2012.}
In this regard research on gravitation belonged to ``Little Science'' until experimental research for the observation of gravitational waves received big funds. \footnote{The concepts ``Little Science'' and ``Big Science''go back to Derek de la Solla Price \cite{SolPri1963}.}\\   

While an institutionalization of the field became achieved, the
situation for young German relativists for entering into research in
the field with a solid financing was not noticeably improved. Much of
{\it research} in general relativity still is done in small groups at
universities in Bremen, Erlangen-Nürnberg, Frankfurt, Hannover, Jena,
Köln, and Oldenburg the continuation of which is not
guaranteed. However, {\it teaching} of general relativity  at
universities  no longer is an exception but has become a normality,
although often as part of elementary particle physics or astrophysics,
and by specialists in these fields. To some extent, Albert Einstein’s
prestige has helped. What has not been achieved is a better standing
of the field in the physics community in terms of more positions at
universities. For the field of gravitation proper, there are even less
positions now than at the beginning of this century. Thus, in 1999, it
again was necessary to sound a warning statement about the decline of
research in gravitation in Germany \cite{MueHe1999}. \\ 

It has been claimed that a ``renaissance'' of general relativity has
occurred in the 1960s to 1970s \cite{Will1989}. While it is true that the
activity in research in general relativity increased in the 1960s, it is
less clear whether one can speak of ``renaissance''. The
research-output before this period was always on a low but continuous 
level; important progress steadily having been made from 1916 to the
1990s.\footnote{Cf. my paper in preparation ``A Golden Age of General 
  Relativity?''} It is also questionable whether the new field of
``Relativistic Astrophysics'' (since the 1950s) and the many papers on
``Cosmology'', should be subsumed under general relativity. As we have seen, in Germany the ``renaissance'' of general relativity started
immediately after world war II and continued slowly, but steadily. 
\section{Acknowledgments}
For specific (oral and written) information I am much obliged to R. Lalli,
Berlin; Don Salisbury, Austin; M. Trümper, Uz\`es; G. Dautcourt,
Berlin, H. Dehnen, Konstanz; D. Hoffmann, Berlin; D.-E. Liebscher,
Potsdam; G. Neugebauer, G. Schäfer, E. Schmutzer all in Jena. Also to
A. Blum who provided Jordan's letter to Heisenberg. For a careful and critical
 reading  of the ms and for sharing his memories with me, I am very
 grateful to Friedrich W. Hehl, Cologne. My thanks go also to Jürgen 
 Renn for his invitation to take part in a research-program at the
 Institute for the History of Science and for his support.  

\end{document}